\documentclass[twocolumn,superscriptaddress,reprint]{revtex4-1}
\usepackage{microtype}
\usepackage{amsmath,amsfonts,amssymb} 

\usepackage{lmodern}
\usepackage{placeins}

\usepackage{graphicx}
\usepackage{color}
\usepackage{xfrac}
\usepackage{epstopdf}
\usepackage{blkarray} 
\usepackage{lipsum}
\usepackage{soul}
\usepackage{multirow}
\usepackage{varioref}
\usepackage[colorlinks=true,
pdfstartview=FitV, 
linkcolor=blue, 
citecolor=blue, 
urlcolor=blue,
linktocpage=true]{hyperref}
\usepackage{cleveref}
\usepackage{soul}

\usepackage[utf8]{inputenc}  
\usepackage{amsmath,braket,amssymb,graphicx,stackrel,color,hyperref}
\usepackage {graphicx}
\usepackage{float}

\usepackage{color}
\definecolor{darkblue}{rgb}{0.,0.,0.5}
\definecolor{darkred}{rgb}{0.5,0.,0.}
\definecolor{darkgreen}{rgb}{0.,0.5,0.}
\hypersetup{colorlinks=true, linkcolor=darkred, citecolor=darkblue, urlcolor=darkblue}

\newcommand{\nnu}{{\boldsymbol{\nu}}}

\begin{document}
	
	\title{Dark vertical conductance of cavity-embedded semiconductor heterostructures}
	\author{Cassia Naudet-Baulieu}
	\author{Nicola Bartolo}%
	\author{Giuliano Orso}
	\author{Cristiano Ciuti}%
	\affiliation{Universit\'{e} de Paris, Laboratoire Mat\'{e}riaux et Ph\'{e}nom\`{e}nes Quantiques, CNRS, 75013 Paris, France}
	
	\begin{abstract}
		We present a linear-response nonlocal theory of the electronic conductance along the vertical (growth) direction of a semiconductor heterostructure embedded in a single-mode electromagnetic resonator in the absence of illumination.
		Our method readily applies to the general class of n-doped semiconductors with parabolic dispersion.
		The conductance depends on the ground-state properties and virtual collective polaritonic excitations that have been determined via a bosonic treatment in the dipole gauge.
		We show that, depending on the system parameters, the cavity vacuum effects can enhance or reduce significantly the dark vertical conductance with respect to the bare heterostructure.
	\end{abstract}
	
	\maketitle

	\section{Introduction}
	
	 Semiconductor heterostructures such as quantum wells (QW) and superlattices are the building blocks of important optoelectronic devices working in the mid and far infrared, such as photodetectors \cite{gendron2004quantum,jagadish2011advances} and quantum cascade lasers \cite{Faist553,Khler2002}.
	In the case of infrared photodetectors, the photo-induced vertical electrical current along the growth direction of the heterostructure is the key detected quantity.
	The sensitivity of a photodetector is limited by the so-called dark current, i.e. the current without illumination \cite{jagadish2011advances}. Several experimental and theoretical studies have investigated such systems \cite{Levine1990,Zussman1991,Liu1993}, pointing out that the main origin of the dark current at low temperatures is the electron tunneling through potential barriers \cite{Levine1993}.
	Recently, an improvement of the performances of Quantum-Well Infrared Photodetectors (QWIP) has been demonstrated for arrays of heterostructures embedded in photonic resonators \cite{palaferri2018room}.
	
	Recent transport experiments in cavity-embedded organic semiconductors \cite{Ebbesen2015,Ebbesen} and in-plane magnetotransport measurements in cavity-embedded high-mobility two-dimensional electron systems \cite{Zurich} have shown that the linear-regime electronic conductance of a material can be significantly affected by a cavity electromagnetic resonator even in the absence of illumination. 
	Early theoretical works exploring the role of a cavity on transport have focused on cavity-modified excitonic linear transport \cite{Pupillo2015,GarciaVidal2015} and charge nonlinear conduction \cite{Hagenmuller2017} for chains of two-level systems in the nonlinear regime where electrons are injected in the excited levels. A recent theoretical work \cite{BartoloPRB18} on in-plane magnetotransport in the presence of Landau electronic levels has shown how virtual polariton excitations can control the charge  transport in the linear regime. Indeed, light-matter interaction can play a pivotal role in determining the electronic transport properties without illumination especially if the strong \cite{Dini2003} or ultra-strong coupling regime \cite{CiutiPRB2005, Anappara2009,Gnter2009,TodorovPRL2010} is achieved. 
	
		In this paper, we report a theoretical study of the electronic linear-regime conductance of a cavity-embedded generic n-doped semiconductor heterostructure with parabolic dispersion, such as n-doped GaAs.
		Our theory considers the electromagnetic-vacuum effects on the conductance: no real photons are injected or created in the resonator.
	 Based on a bosonized Hamiltonian, we calculate the vertical conductance in the growth direction within a nonlocal linear-response Kubo approach \cite{FlensbergBook}. In order to describe the collective electronic effects, we consider the system Hamiltonian in the dipole gauge \cite{DipolarGaugeRoy,TodorovPRB,DeBernardis2018,Cortese:19} where two interaction terms emerge, describing respectively the so-called depolarization shift and the light-matter coupling. The present general framework is applied to quantum well heterostructures.
	Numerical results showing how the dark conductance is controlled by the cavity without illumination are presented.
	
	The article is organized as follows. In Sec.~\ref{sec:model}, we introduce the general theoretical bosonic framework to determine collective light-matter excitations. 
	In Sec.~\ref{sec:GeneralConductivity}, within a many-body nonlocal Kubo formalism  we provide the general expression for the dark vertical conductance for n-doped heterostructures depending on the properties of the ground state and of the polariton excitations.
	Numerical results for specific heterostructures are presented and discussed in Sec.~\ref{sec:Results}.
	Finally, we draw our conclusions and perspectives in Sec.~\ref{sec:conclusions}. The most technical details about the theoretical model are reported in Appendix \ref{App:Hopfield} and  \ref{App:Kubo}.

	\section{Hamiltonian framework and collective light-matter states}\label{sec:model}
	
\begin{figure}
		\includegraphics[width=0.8\linewidth]{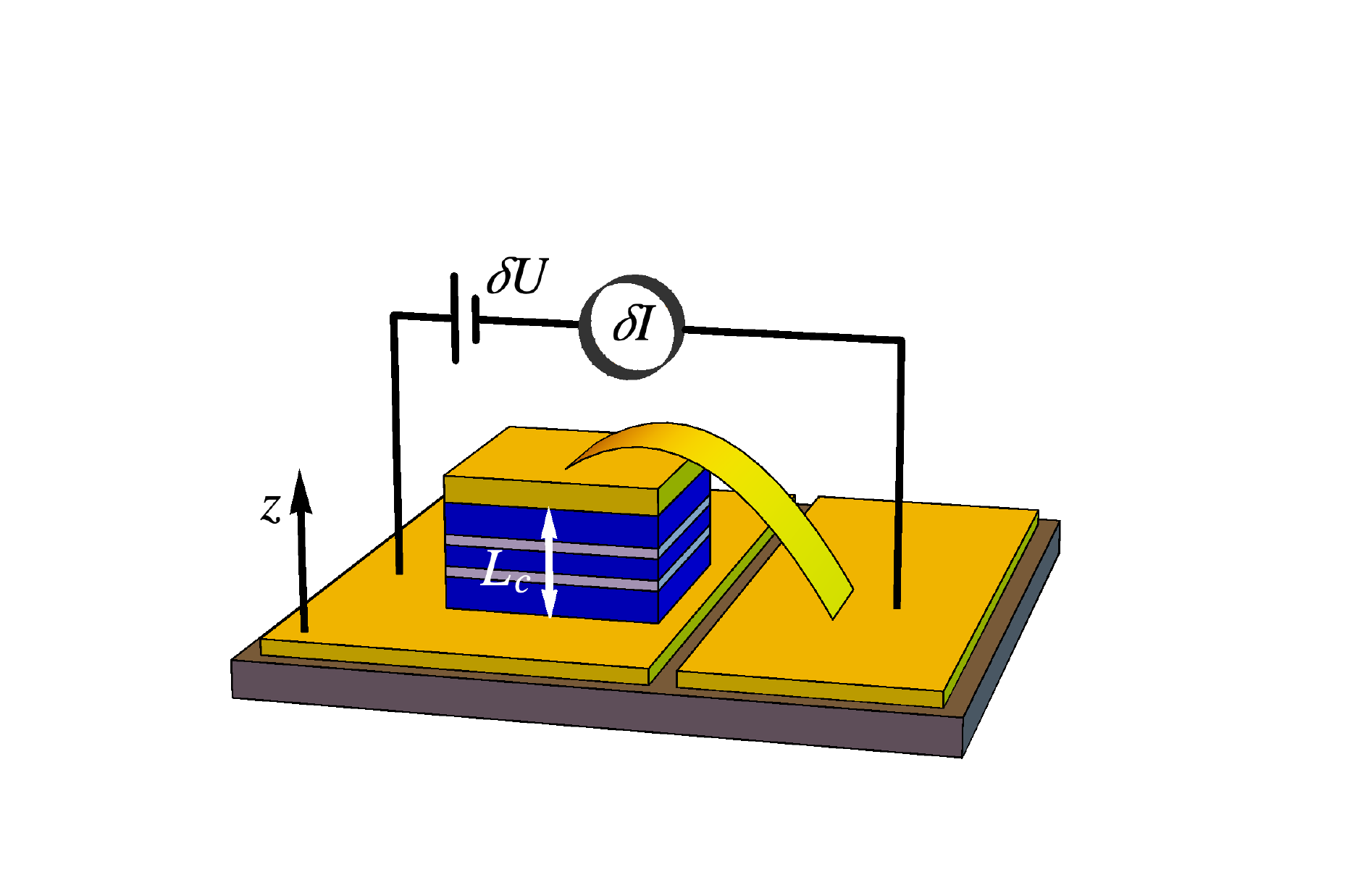}
		\caption{\label{fig:System}
			Sketch of the considered system, namely a n-doped semiconductor heterostructure grown along the $z$ direction and embedded in a single-mode photonic cavity of length $L_c$. The specific LC electromagnetic resonator here depicted is of the same type as in Ref. \cite{Paulillo2017}, where it is possible to tune the frequency of the cavity mode by changing the inductive bridge without altering the capacitive geometry of the cavity-embedded heterostructure.
			The key quantity analyzed in the present work is the vertical dark conductance $G =\delta I /\delta U$, where $\delta I$ is the dark current (no illumination) and $\delta U$ the voltage applied between the two leads separated by the cavity spacer length $L_{c}$.
		}
	\end{figure}

\begin{figure}
	\includegraphics[width=0.95\linewidth]{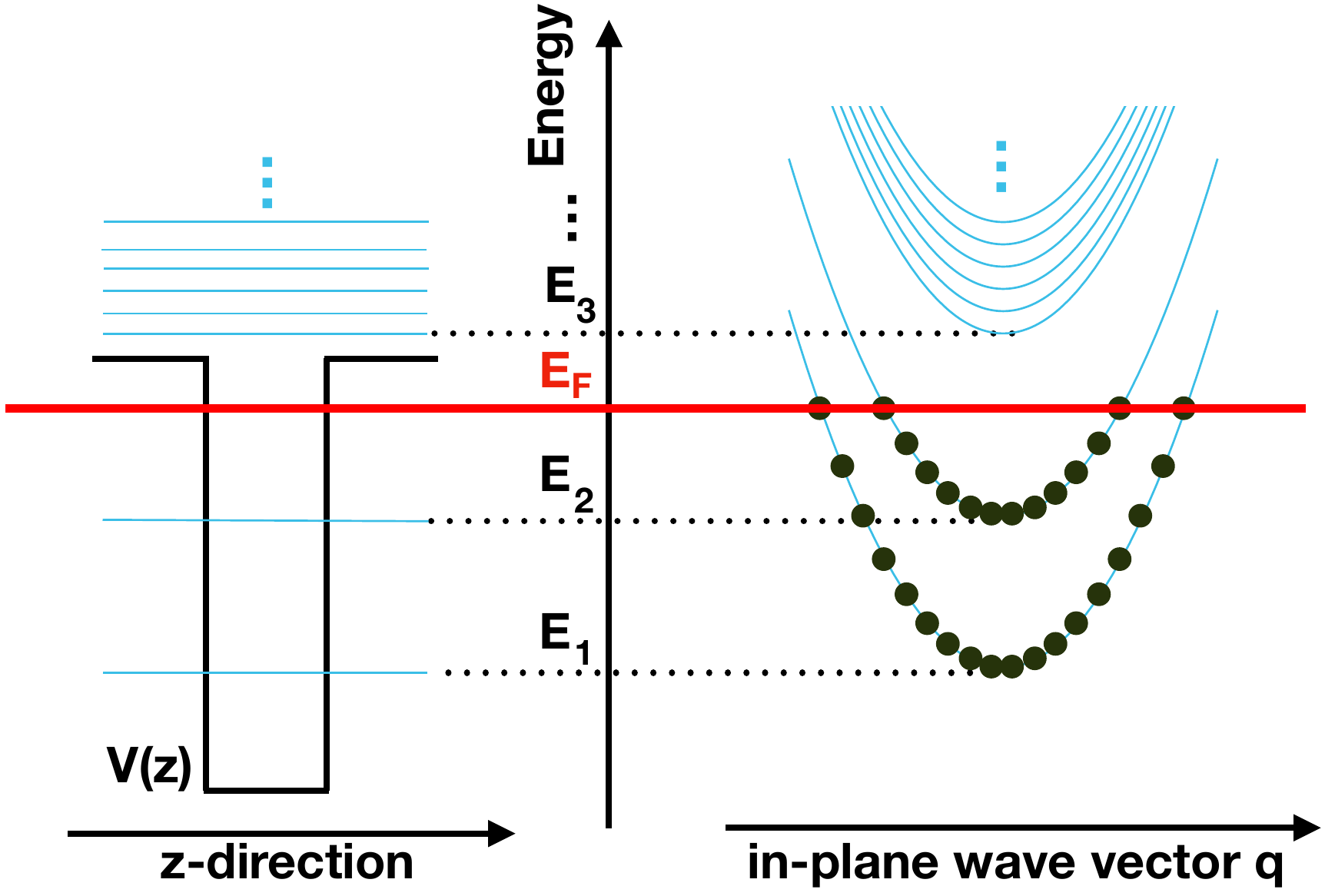}
	\caption{\label{fig:Band}
		Left panel: example of potential $V(z)$ of a quantum well structure. The blue horizontal lines mark the energies of the one-electron eigenstates obtained by solving Eq.~\eqref{Eq:Schrodinger}. Note that our formalism can be applied to an arbitrary potential shape $V(z)$.
		Right panel: parabolic dispersion of the corresponding conduction subband energies as a function of the in-plane wavector. The red horizontal line depicts the Fermi energy level $E_F$.
		In this example, only two subbands are populated, corresponding to $j_F=2$ in our notation.
	}
\end{figure}

	The system considered here is an arbitrary n-doped semiconductor heterostructure with small non-parabolicity epitaxially grown along the $z$ direction and embedded in a single-mode electromagnetic cavity.
	Single-mode electromagnetic resonators with tunable frequency can be obtained, for example, via LC-like structures \cite{lee2010review,TodorovPRX,Paulillo2014,Paulillo2017} (see Fig.~\ref{fig:System}). We assume the electrons to be free in the transverse direction where $S=L_x\times L_y$ is the transverse area.
	The single-particle electronic eigenfunctions can be written as
	\begin{equation}\label{Eq:WaveFunctions}
	\varphi_{j,{\bf q}}(\boldsymbol{r})=\frac{e^{i {\bf q}\cdot\boldsymbol{r}_\parallel}}{\sqrt{S}}\phi_j(z),
	\end{equation}
	where $\boldsymbol{r}_\parallel=\{x,y\}$ is the in-plane electron position and ${\bf q}=\{k_x,k_y\}$ the in-plane wavevector. 
	The eigenfunctions $\phi_j$ and their energies $E_j$, are found by solving the one-dimensional Schr\"odinger equation for the motion along $z$.
	Within the effective mass approximation \cite{BookKlingshirn}, the latter reads
	\footnote{For simplicity, we neglect here the variation of effective mass along the heterostructure.}:
	\begin{equation}\label{Eq:Schrodinger}
	\left(-\frac{\hbar^2}{2m_\star}\frac{\partial^2\,}{\partial z^2}+V(z)\right)\phi_j(z)=E_j\phi_j(z)
	\end{equation}
	with $m_\star$ the effective mass and $V(z)$ the potential describing an heterostructure of arbitrary shape.
	The in-plane energy dispersion of the state $\varphi_{j,{\bf q}}$ is $E_{j,{\bf q}}=E_j+\hbar^2q^2/(2m_\star)$, as sketched in  Fig.~\ref{fig:Band}.
	In the second quantization framework, it is convenient to introduce the fermionic operators $\hat{c}_{j,{\bf q}}$  ($\hat{c}_{j,{\bf q}}^\dagger$) which annihilate (create) an electron in the $j$-th band with wave-vector ${\bf q}$.
	In the absence of interactions,  the many-body ground state is a Fermi sea with only the first $j_F$  subbands populated (cf. Fig.~\ref{fig:Band}), namely: 
	\begin{equation}
	\ket{\rm FS}=\prod_{j\leq j_F}\prod_{k<k_{Fj}}\hat{c}_{j,{\bf k}}^\dagger\, \ket{\rm vacuum},
	\label{eq:defFS}
	\end{equation}
	where $k=|{\bf k}|$, $k_{Fj}$ is the Fermi wave vector associated with the $j$-th band, while $\ket{\rm vacuum}$ is the electronic and photonic vacuum.
	To describe the light-matter interaction for the heterostructure it is convenient to introduce collective electronic excitations \cite{CiutiPRB2005,TodorovPRB,Cortese:19} described by the operators
	\begin{equation}\label{Eq:ColectiveExcitations}
	\hat{b}_{l,j}^\dagger=\frac{1}{\sqrt{N_{l,j}}}\sum_{{\bf q}}\hat{c}_{l,{\bf q}}^\dagger\hat{c}_{j,{\bf q}},
	\end{equation}
	for $j\leq j_F$ and $j<l$.
	The number $N_{l,j}=N_j-N_l>0$ is the difference between the occupation numbers in the two conduction subbands and ensures the normalization of the state $\hat{b}^\dagger_{l,j} \vert {\rm FS} \rangle$.
	To simplify the notation, we compact the double index $l,j$ into $\nnu=\{l,j\}$, assuming that any sum on such index runs over all the index pairs satisfying $l>j$ and $j\leq j_F$, unless differently specified.
	Note that in the limit where $N_\nnu\gg1$ the $\hat{b}_\nnu$ operators satisfy bosonic commutation relations, that is $ [b_\nnu,\, b_{\nnu'}^\dagger]=\delta_{\nnu\nnu'}$  \cite{DeLiberatoPRL09}
\footnote{More specifically, $\left[\hat{b}_{l,j},\hat{b}_{l,j}^\dagger\right]=\frac{1}{N_{l,j}}\left[\hat{N}_j-\hat{N}_l\right]$ with $\hat{N}_j=\sum_{\bf q}\hat{c}_{j,{\bf q}}^\dagger\hat{c}_{j,{\bf q}}$ the population operator in subband $j$ and $N_{l,j} = N_j-N_l\gg 1$. As long as one considers states with a total number of excitations much smaller than $N_{l,j}$, one has \cite{DeLiberatoPRL09} $\frac{1}{N_{l,j}}\left[<\hat{N}_j>-<\hat{N}_l>  \right]\simeq 1$.

This implies that within a few-excitation subspace one can consider $\left[\hat{b}_{l,j},\hat{b}_{l,j}^\dagger\right] = 1$.}.

The total Hamiltonian of the considered system can be arranged as the sum of four terms:
\begin{equation}\label{Eq:TotalHamiltonian}
\hat{\mathcal{H}}_0=\hat{\mathcal{H}}_e+\hat{\mathcal{H}}_c+\hat{\mathcal{H}}_{LM}+\hat{\mathcal{H}}_{dep},
\end{equation}
where we will neglect all constant contributions.

The noninteracting electronic contribution $\hat{\mathcal{H}}_e$ can be written in terms of the operators introduced in Eq.~\eqref{Eq:ColectiveExcitations} and of the transition energies $\hbar\omega_{\nnu}=E_l-E_j$ \footnote{This can be done by projecting the full electronic Hamiltonian over the manybody excited states $\hat{b}_\nnu^\dagger\ket{\rm FS}$ \cite{TodorovPRB,DeLiberatoPhD}}:
	\begin{equation}
	\hat{\mathcal{H}}_e=\hbar \sum_\nnu \omega_\nnu\, \hat{b}_\nnu^\dagger\, \hat{b}_\nnu.
	\label{electronic_hamiltonian} 
	\end{equation}
	
	In the following, we wish to consider the Hamiltonian for an arbitrary semiconductor heterostructure embedded in a single-mode photonic cavity.
	Calling $\omega_c$ the frequency of the photon mode, the bare cavity contribution to the system Hamiltonian is given by
	\begin{equation}
	\hat{\mathcal{H}}_c=\hbar\omega_c\hat{a}^\dagger\hat{a},
	\end{equation}
	where $\hat{a}$  ($\hat{a}^\dagger$) is the bosonic annihilation (creation) operator for a cavity photon.
	
	To describe electromagnetic interactions, we will work here in the dipolar gauge \cite{DipolarGaugeRoy,TodorovPRB}, giving rise to two interaction terms.
	The first describe the coupling between the cavity photon mode and the collective intersubband excitations:
	\begin{equation}
	\hat{\mathcal{H}}_{LM}=i \hbar
	\left(\hat{a}^\dagger-\hat{a}\right)
	\sum_\nnu \Omega_\nnu
	\left(\hat{b}_\nnu^\dagger+\hat{b}_\nnu\right)\,,
	\label{eqLMhamiltonian}
	\end{equation}
	where the collective vacuum Rabi frequency $\Omega_\nnu$ reads \cite{TodorovPRB}
	\begin{equation}
	\Omega_{\nnu}=\sqrt{\frac{\hbar N_{\nnu} e^2}{8\epsilon_0\epsilon_r m_\star^2 S}}\sqrt{\frac{\omega_c}{L_c}}\frac{\int dz\,\xi_{\nnu}(z)}{\omega_{\nnu}}
	\label{Eq:Omega}
	\end{equation}
	with $L_c$ the cavity length, $\epsilon_0\epsilon_r$ the dielectric constant of the material filling the cavity, and
	\begin{equation}\label{Eq:DefXi}
	\xi_{\nnu}(z)=[\partial_z\phi_{l}(z)]\phi_{j}(z)-\phi_{l}(z)[\partial_z\phi_{j}(z)].
	\end{equation}
	A relevant quantity indicating the strength of the light-matter coupling of a given transition is the resonant Rabi coupling
	$\Omega_\nnu^{\rm res}=\Omega_\nnu|_{\omega_c=\omega_\nnu}$.
	Note that the light-matter interaction Hamiltonian~\eqref{eqLMhamiltonian} includes both resonant and anti-resonant (counter-rotating wave) terms.
	The second interaction term is called depolarization shift Hamiltonian, which describes the self-interaction of the electronic polarization.
	In the considered system, it reads \cite{TodorovPRB}
	\begin{equation}
	\mathcal{H}_{dep}=\hbar\sum_{\nnu}\sum_{\nnu'}\Xi_{\nnu}^{\nnu'}
	\left(\hat{b}_\nnu^\dagger+ \hat{b}_\nnu\right)
	\left( \hat{b}_{\nnu'}^\dagger+ \hat{b}_{\nnu'}\right),
	\label{eq:dep_shift}
	\end{equation}
	where
	\begin{equation}
	\label{Eq:collectivedepshift}
	\Xi_{\nnu}^{\nnu'}=\frac{\hbar\, e^2}{8\epsilon_0\epsilon_rm_\star^2 S}\frac{\sqrt{N_{\nnu}N_{\nnu'}}}{\omega_{\nnu}\omega_{\nnu'}}\int dz\, \xi_{\nnu}(z)\xi_{\nnu'}(z).
	\end{equation}
	The depolarization Hamiltonian $\mathcal{H}_{dep}$ is due to the collective part of the Coulomb interaction \cite{LeePRB99,DeLiberatoPRB12}.
	The other Coulomb terms are responsible for non-collective electron-electron scattering which will be neglected here. 
	

	In this work, we will focus on the case where all the subbands $j\leq j_F$  are macroscopically occupied. 
	In this limit, the collective excitations behave as bosons and the Hamiltonian~\eqref{Eq:TotalHamiltonian} is quadratic in the bosonic operators $\hat{a}$ and $\hat{b}_\nnu$.
	Hence, the eigenstates of the system can be determined by performing a Hopfield-Bogoliubov transformation \cite{Hopfield,CiutiPRB2005,nataf2010no,DeLiberatoPRB12}.
	The Hamiltonian can be diagonalized by introducing the bosonic hybrid light-matter polaritonic annihilation operators
	\begin{equation}\label{eq:def_pr}
	\hat{p}_r=w_r \hat{a}+ y_r \hat{a}^\dagger +\sum_\nnu \left( x_{r,\nnu} \hat{b}_\nnu + z_{r,\nnu} \hat{b}_\nnu^\dagger \right).
	\end{equation}
	To ensure the bosonic commutation relations $\left[\hat{p}_r,\hat{p}_{r'}^\dagger\right]=\delta_{r,r'}$,
	the coefficients must satisfy the hyperbolic normalization
	\begin{equation}
	|w_r|^2-|y_r|^2 +\sum_{\nnu} \left( |x_{r,\nnu}|^2-|z_{r,\nnu}|^2  \right) =1.
	\end{equation}
	In terms of such polariton operators, the total Hamiltonian~\eqref{Eq:TotalHamiltonian} reads
	\begin{equation}
	\hat{\mathcal{H}}_0=\hbar \sum_r \omega_r^I\, \hat{p}_r^\dagger \hat{p}_r.
	\end{equation}
	The Hopfield-Bogoliubov coefficients and the polariton frequencies $\omega_{r}^I$ are determined by solving the eigenvalue equation ${\bf M} \vec{v}_r = \omega_{r}^I \vec{v}_r $, where the vector
	$\vec{v}_r = \left(w_r, x_{r,\nnu_1}, x_{r,\nnu_2}, \cdots, y_r, z_{r,\nnu_1}, z_{r,\nnu_2}, \cdots \right)^{T}$
	and the matrix ${\bf M}$ is given explicitly in Appendix \ref{App:Hopfield}, Eq.~\eqref{Eq:App:M}.

	\section{Dark vertical Conductance}\label{sec:GeneralConductivity}
	
	Our goal in this section is to derive an expression for the dark vertical conductance of the cavity-embedded heterostructure, i.e. the linear-regime electronic transport along the growth direction without illumination.
	In order to do so, we first determine the nonlocal conductivity via a linear-response Kubo approach~\cite{FlensbergBook,Girvin2019}.
	Specifically, we are interested in the linear response when a small bias voltage $\delta U$ is applied along the vertical (growth) direction $z$ between the two leads at the edges of the sample separated by the cavity spacer length $L_{c}$ [cf. Fig.~\ref{fig:System}].  
	
	Assuming that the heterostructure is translationally invariant in the $xy$-plane, but not along the growth direction $z$, the electric field within the sample depends on $z$ but not on ${\bf r}_\parallel$.
	This allows us to simplify the general Kubo expression by integrating out the transverse directions, as detailed in Appendix~\ref{App:Kubo}.
	The field in the $z$ direction and the voltage are related by $E_z(z) = -\partial_z U(z)$ with
	
	\begin{equation}
	\delta U = - \int_{-L_{c}/2}^{L_{c}/2} dz E_z(z) = U(L_{c}/2)-U(-L_{c}/2).
	\end{equation}

	The variation of current density in the low-temperature limit is determined through the non-local response function $\chi(z,z')$ via
	\begin{equation}\label{Eq:DefCurrentVariation}
	\delta\langle\hat{J}_z\rangle(z)=\int_{-L_{c}/2}^{L_{c}/2} dz'\, \chi(z,z')\,E_z(z'),
	\end{equation}
	with
	\begin{align}\label{Eq:DefChi}
	\chi(z,z')=&2\hbar S\sum_{b>1}\frac{\eta_{b}}{{\mathcal E}_{b}-{\mathcal E}_{1}}
	\frac{
		\braket{\Psi_1|\hat{J}^\nabla_z(z)|\Psi_b}
		\!\!
		\braket{\Psi_b|\hat{J}^\nabla_z(z')|\Psi_1}}{{\left({\mathcal E}_{b}-{\mathcal E}_{1}\right)^2+\eta_{b}^2}},
	\end{align}
	where $\{\ket{\Psi_b}\}_{b=1,2,\dots}$ are the many-body ground ($b=1$) and excited ($b>1$) states of the system considered, being ${\mathcal E}_{b}$ the corresponding energies.
	The quantity $\eta_{b}=\hbar/\tau_{b}$ depends on the phenomenological scattering time $\tau_{b}$ \cite{FlensbergBook,AllenChapter}.
	For purely electronic excitations, $\tau_{b}$ corresponds to the Drude transport scattering time $\tau_0$.
	Finally, the operator $\hat{J}_z^\nabla(z)$ is obtained from the $z$-component of the paramagnetic current density operator, upon integration on the transverse plane:
	\begin{align}\label{Eq:CurrentOperatorReduced}
	\hat{J}_z^\nabla( z)&=i \frac{ e\hbar}{2m_\star S}\sum_\nnu \sqrt{N_\nnu}\, \xi_\nnu(z) \left(\hat{b}_\nnu - \hat{b}^\dagger_\nnu\right).
	\end{align}


Note that the inhomogeneity of the semiconductor heterostructure implies that the induced current $ \delta \langle\hat{J}_z\rangle$ actually depends on $z$ [cf. Eq.~\eqref{Eq:DefCurrentVariation}].
As a result, applying an infinitesimal bias $\delta U$ between the leads at $z = \pm L_c/2$, implies a $z$-dependent change $\delta \langle\hat{\rho}_z\rangle(z)$ of the electronic charge density
where $\int_{-L_c/2}^{L_c/2} \delta \rho(z) dz = 0$. In other words, while the total charge is conserved in the semiconductor heterostructure, there are local variations of the charge density.
This explains why $\delta \langle\hat{J}_z\rangle$ is not constant along the semiconductor heterostructure. As an example, it can be analytically proven from the general linear response Kubo expression applied to $\delta \langle\hat{\rho}_z\rangle(z)$ and $ \delta \langle\hat{J}_z\rangle$ for the case of a noninteracting system, that the following continuity expression holds
\begin{equation}\label{Eqref:continuityequation}
\frac{\partial \,\delta \langle\hat{J}_z\rangle(z)}{\partial\, z} = - \frac{\delta \langle\hat{\rho}_z\rangle(z)}{\tau_0}, 
\end{equation}
where $\tau_0 = 1/\eta_0$ is the inelastic scattering time (the same for all subbands). Indeed, this expression shows the link between the induced current  and the dissipation through charge accumulations. In the case of an interacting system, an analogous continuity equation holds with a more complex form.

The derivation of the conductance from the two-point response function $\chi$ in non-translationally-invariant systems is a subtle task which has been addressed in detail for elastic scattering in disordered systems \cite{Stone_1989, Kane1988}.
The application of a bias $\delta U$ across the sample induces a current $\delta I$ coming out of the lead at $z=L_{c}/2$ (Fig.~\ref{fig:System}).
The latter is related to the current density~\eqref{Eq:DefCurrentVariation} by $\delta I=S \delta\langle\hat{J}_z\rangle(L_{c}/2)$.

Assuming a homogeneous flow in the metallic contacts, such a $\delta I$ must be also equal to the current entering through the lead at $z=-L_{c}/2$, that is, $\delta I=S \delta\langle\hat{J}_z\rangle(-L_{c}/2)$.
To ensure this, we impose periodic boundary conditions to the functions $\phi_j$ when solving the Schr\"odinger equation~\eqref{Eq:Schrodinger}.
With this choice, definitions~\eqref{Eq:DefXi}, \eqref{Eq:DefChi}, and~\eqref{Eq:CurrentOperatorReduced} imply $\hat{J}^\nabla_z(-L_c/2)=\hat{J}^\nabla_z(L_c/2)$, which in turns ensure the periodicity of the current at the leads.
	
Qualitatively, we can assume the variation of the electric field to be negligible, so that the current reads: 
\begin{equation}\label{Eq:IPP}
\delta I=S\, E_z\int_{-L_c/2}^{L_c/2}\!\! dz' \chi(L_c/2,z')
\end{equation}
from which the general expression for the two-probe conductance is 
\begin{equation}
G=\frac{\delta I}{\delta U}=-\frac{S}{L_c}\,\int_{-L_c/2}^{L_c/2}\!\! dz' \chi(L_c/2,z').
\end{equation}
	
	\subsection{Conductance for non-interacting electrons}\label{subsub:nonInt}
	Let us first consider the behavior of a non-interacting system (no light-matter interaction, no depolarization shift, i.e. $\hat{\mathcal{H}}_{LM}=\hat{\mathcal{H}}_{dep}=0$).
	In this case $\ket{\Psi_1}=\ket{\rm FS}$ and the only excited states connected to the ground state via $\hat{J}^\nabla_z(z)$ are  $\hat{b}_\nnu^\dagger\ket{\rm FS}$, which are states containing a collective electron-hole excitation above the Fermi sea. This simplifies Eq.~\eqref{Eq:DefChi} to the form
	\begin{equation}\label{eq:sigmaNonInt}
	\chi_{_{\!N\!I}}(z,z')=\frac{\hbar\, n_e e^2}{2 m_\star^2} \sum_{\nnu} \frac{\tau_0}{\omega_{\nnu}}\,
	\frac{ \widetilde{\xi}_{\nnu}(z) \widetilde{\xi}_{\nnu}(z')}{1+\left(\tau_0\omega_{\nnu}\right)^2},
	\end{equation}
	where $\widetilde{\xi}_{\nnu}(z)=\sqrt{N_\nnu/N_e}\,{\xi}_{\nnu}(z)$.
	The noninteracting conductance reads 
	\begin{equation}\label{eq:GNonInt}
	G_{\!N\!I}=-\frac{S}{L_c}\frac{\hbar\, n_e e^2}{2m_\star^2} \sum_{\nnu} \frac{\tau_0}{\omega_{\nnu}}\,
	\frac{ \widetilde{\xi}_{\nnu}(L_c/2)\int_{-L_c/2}^{L_c/2}\widetilde{\xi}_{\nnu}(z')dz'}{1+\left(\tau_0\omega_{\nnu}\right)^2}.
	\end{equation}
	
	\subsection{The cavity case: \\conductance in the absence of illumination}\label{subsub:Int}
	In presence of interaction with the cavity quantum field, the manybody ground state of the system differs from the noninteracting Fermi sea, that is, $\ket{\rm GS} \neq \ket{\rm FS}$. The cavity-dressed polaritonic excited states are defined by $\hat{p}_r^\dagger\ket{\rm GS}$.
	The current operator~\eqref{Eq:CurrentOperatorReduced} can be suitably rewritten exploiting the inverse of Eq.~\eqref{eq:def_pr}:
	$\hat{b}_\nnu=\sum_r \left( x_{r,\nnu}^* \hat{p}_r - z_{r,\nnu} \hat{p}^\dagger_r \right)$.
	Thus, equation~\eqref{Eq:DefChi} can be easily applied to the new set of dressed manybody states to get
	\begin{equation}\label{eq:sigmaInt}
	\chi(z,z')=\frac{\hbar\, n_e e^2 }{2m_\star^2 } \sum_{r}\frac{\tau_r}{\omega_{r}^I}\,
	\frac{ {\widetilde{\xi}_{r}^{\rm eff^*}}\!(z)\,\widetilde{\xi}_{r}^{\rm eff}(z')}{1+\left(\tau_r\omega_{r}^I\right)^2}
	\end{equation}
	with 
	\begin{equation}
	\widetilde{\xi}_{r}^{\rm eff}(z)=\sum_{\nnu}
	\left(x_{r,\nnu}+z_{r,\nnu}\right)\,\widetilde{\xi}_{\nnu}(z).
	\end{equation} 
	This finally gives the dark vertical conductance 
	
	\begin{equation}\label{eq:GInt}
	G=-\frac{S}{L_c}\frac{\hbar \,n_e e^2}{2m_\star^2 } \sum_{r}\frac{\tau_r}{\omega_{r}^I}\,
	\frac{ \widetilde{\xi}_{r}^{\rm eff^*}(L_{c}/2)\int_{-L_c/2}^{L_c/2}\widetilde{\xi}_{r}^{\rm eff}(z')dz' }{1+\left(\tau_r\omega_{r}^I\right)^2}.
	\end{equation}
	
	Note that several effects contribute in determining how the cavity dark conductance  \eqref{eq:GInt} differs from the non-interacting one \eqref{eq:GNonInt}. First, the electronic states are dressed by photonic ones due to  the interactions, which can strongly modify the spatial localization of the manybody wavefunctions. This determines the spatial shape of $\widetilde{\xi}_r^{\rm eff}$ and is responsible for what we call the orbital renormalization of the conductance.
	Second, the interactions change the energetic cost associated with a polaritonic transition, thus altering the conductance.
	The third effect is the modification of the scattering times $\tau_r$, which are associated with excited states having a mixed light-matter nature \cite{BartoloPRB18}. These hybrid scattering times are a weighted combination of the Drude electronic scattering time $\tau_0$ and of a photonic transport scattering time $\tau_p$:
	\begin{equation}\label{Eq:ScatteringTimesMix}
	\frac{1}{\tau_r}=\frac{W_{e,r}}{\tau_0}+\frac{1-W_{e,r}}{\tau_p},
	\end{equation}
	where the electronic weight of the polariton created by $p_r^\dagger$ is 
	\begin{equation}
	W_{e,r}=\sum_{\nnu}|x_{r,\nnu}|^2-\sum_{\nnu}|z_{r,\nnu}|^2.
	\end{equation}
	Note that $\tau_p$ is not the photon lifetime inside the resonator, but a transport scattering time expected to be much longer \cite{BartoloPRB18}.
	In what follows, we will consider $\tau_p\gg\tau_0$, implying \mbox{$\tau_r=\tau_0/W_{e,r}$}.
	
	\subsection{Two-subband approximation}\label{sec:MainTransition}
	
	In general, there is a subtle interplay between orbital, energetic and scattering-time effects determining the conductance of a cavity-embedded heterostructure.
	One can get some analytical insight in the special case in which one transition gives a dominant contribution to the conductance.
	This is the case, for instance, when one transition frequency is significantly smaller than the others, since high-frequency terms are quickly suppressed in the sums of Eqs.~\eqref{eq:GNonInt} and~\eqref{eq:GInt}.
	In this context, the four-by-four Hopfield-Bogoliubov matrix~\eqref{Eq:App:Mreduced} can be analytically diagonalized.
	
	Let $\bar{\nnu}$ be the two-component index associated with the dominant transition, the noninteracting conductance reads
		\begin{equation}
	G_{\!N\!I,\bar{\nnu}}=-\frac{S}{L_c}\frac{\hbar\, n_e e^2}{2m_\star^2}  \frac{\tau_0}{\omega_{\bar{\nnu}}}\,
	\frac{ \widetilde{\xi}_{\bar{\nnu}}(L_c/2)\int_{-L_c/2}^{L_c/2}\widetilde{\xi}_{\bar{\nnu}}(z')dz'}{1+\left(\tau_0\omega_{\bar{\nnu}}\right)^2}.
	\end{equation}

Making use of the exact relation holding for Hopfield coefficients \cite{BartoloPRB18}
	\begin{equation}
	\left| x_{r,\bar{\nnu}}+z_{r,\bar{\nnu}} \right|^2 = W_{e,r}\, \frac{\omega_{r}^I}{ \omega_{\bar{\nnu}}},
	\end{equation}
	the cavity dark conductance can be recast as
	\begin{equation}
	G_{\bar{\nnu}}=G_{\!N\!I,\bar{\nnu}}
	\sum_{r=LP,UP}
	\frac{ 1+\left(\tau_0\omega_{\bar{\nnu}}\right)^2 }{1+\left(\tau_r \omega_{r}^I\right)^2}.
	\end{equation}
	Note that in this one-transition approximation there are only two polaritonic branches: the lower ($r=LP$) and upper ($r=UP$) one.
	
	Two relevant limits can be addressed.
	First, let us consider the limit $\omega_c \to 0$, where only the depolarization shift effect matters.
	In this case, we get
	\begin{align}\label{Eq:2SBA0}
	G_{\omega_c\to 0,\bar{\nnu}}&= G_{\!N\!I,\bar{\nnu}} \frac{ 1+\left(\tau_0\omega_{\bar{\nnu}}\right)^2 }{1+\left(\tau_0 \omega_{\bar{\nnu}}\right)^2  \left( 1+ 4 \frac{\Xi_{\bar{\nnu}}^{\bar{\nnu}}}{\omega_{\bar{\nnu}}} \right) }
	\nonumber\\
	&\xrightarrow{\tau_0 \omega_{\bar{\nnu}} \gg 1}
	\frac{G_{\!N\!I,\bar{\nnu}}}{1+ 4 \frac{\Xi_{\bar{\nnu}}^{\bar{\nnu}}}{\omega_{\bar{\nnu}}}}\leq G_{\!N\!I,\bar{\nnu}}
	\nonumber\\
	&\xrightarrow{\tau_0 \omega_{\bar{\nnu}} \ll 1} G_{\!N\!I,\bar{\nnu}}.
	\end{align}
	The second limit is that of a high-frequency resonator: $\omega_c \to \infty$.
	The asymptotic value of the conductance reads
	\begin{align}\label{Eq:2SBAInf}
	G_{\omega_c\to\infty,\bar{\nnu}}&= G_{\!N\!I,\bar{\nnu}} \frac{ 1+\left(\tau_0\omega_{\bar{\nnu}}\right)^2 }{1+\left(\tau_0 \omega_{\bar{\nnu}}\right)^2  \left[ 1+ 4 
		\left[\frac{\Xi_{\bar{\nnu}}^{\bar{\nnu}}}{\omega_{\bar{\nnu}}}
		-\left(\frac{\Omega_{\bar{\nnu}}^{\rm res}}{\omega_{\bar{\nnu}}}\right)^2 \right] \right] }
	\nonumber\\
	&\xrightarrow{\tau_0 \omega_{\bar{\nnu}} \gg 1}
	\frac{G_{\!N\!I,\bar{\nnu}}}{ 1+ 4 
		\left[\frac{\Xi_{\bar{\nnu}}^{\bar{\nnu}}}{\omega_{\bar{\nnu}}}
		-\left(\frac{\Omega_{\bar{\nnu}}^{\rm res}}{\omega_{\bar{\nnu}}}\right)^2 \right] } \leq G_{\!N\!I,\bar{\nnu}}
	\nonumber\\
	&\xrightarrow{\tau_0 \omega_{\bar{\nnu}} \ll 1} G_{\!N\!I,\bar{\nnu}}.
	\end{align}
	Note that one always has  $\frac{\Xi_{\bar{\nnu}}^{\bar{\nnu}}}{\omega_{\bar{\nnu}}}
	\geq \left(\frac{\Omega_{\bar{\nnu}}^{\rm res}}{\omega_{\bar{\nnu}}}\right)^2$ due to the Cauchy-Schwartz inequality
	\begin{equation}
	\int_{-L_c/2}^{L_c/2} dz\,\xi_{\bar{\nnu}}^2(z) \geq
	\frac{1}{L_c} \left[ \int_{-L_c/2}^{L_c/2} dz\,\xi_{\bar{\nnu}}(z)\right]^2.
	\end{equation}
	In the considered limits, since there is no scattering-time hybridization effect, the conductance modification is solely due to orbital and energetic effects. We see that the interactions always reduce the conductance with respect to the noninteracting case in the considered limits.
	It is also interesting to note that 
	\begin{equation}\label{Eq:Inequalities}
	G_{\omega_c\to0,\bar{\nnu}}\leq G_{\omega_c\to\infty,\bar{\nnu}}\leq G_{\!N\!I,\bar{\nnu}}.
	\end{equation}
	
	
	\section{Numerical results for quantum well structures}\label{sec:Results}
	
	In order to understand the impact of the cavity on the physical properties of the embedded heterostructure,  we consider some paradigmatic examples.
	We study both the case of a single quantum well and of a multiple quantum well heterostructure.
	For all the numerical results presented in this paper, we carefully verified that convergence was reached by increasing the number of intersubband transitions up to a large enough value.

	\subsection{Single Quantum Well}

	\begin{figure}
	\includegraphics[width=0.95\linewidth]{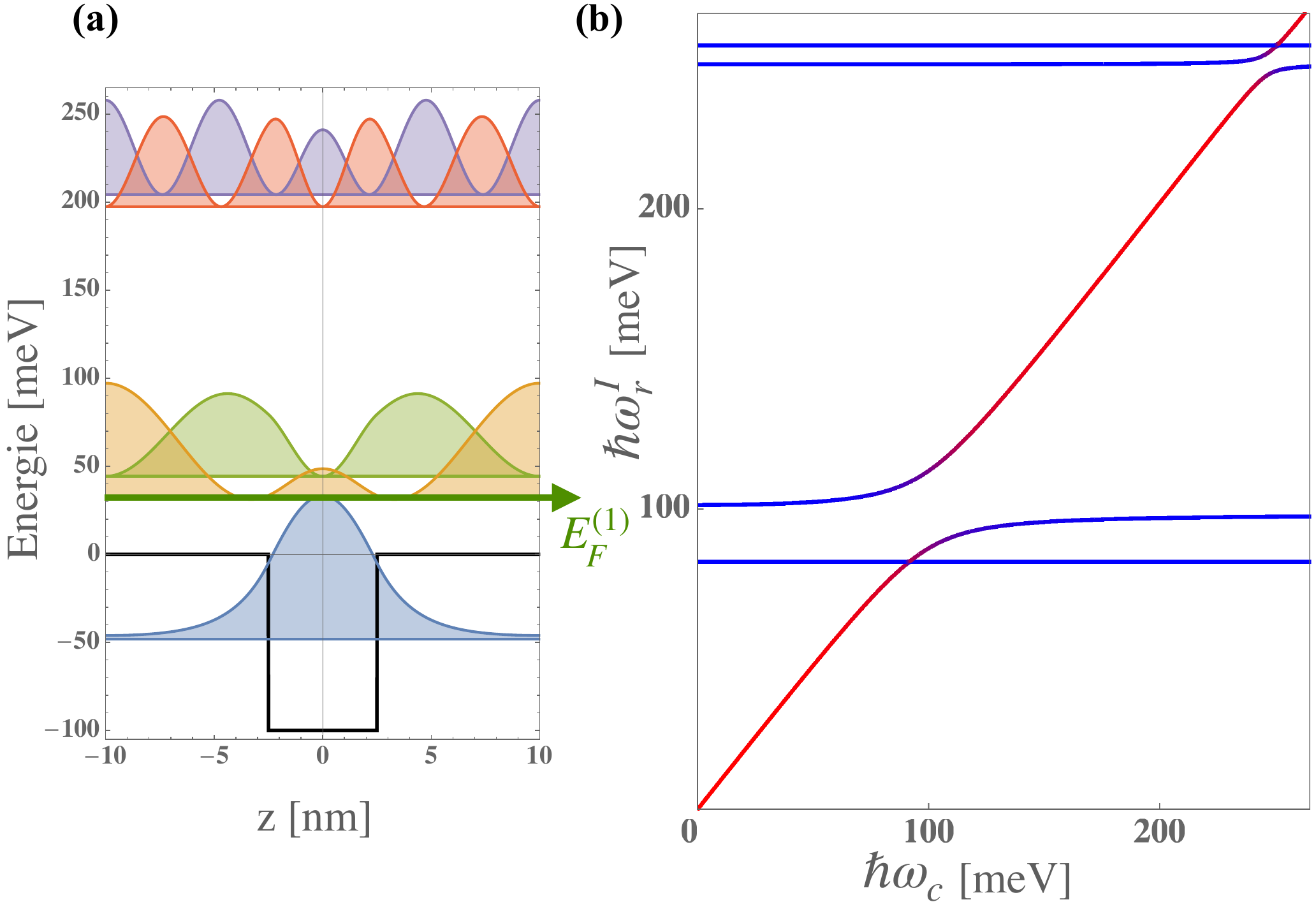}
	\caption{\label{fig:OneWell}
		Panel~(a): single-electron energy levels and wavefunctions (square modulus, arbitrary units, offset vertically for clarity) for a cavity-embedded quantum well.
		System parameters: $V_0=100\, {\rm meV}$, $L_{QW}=5\,{\rm nm}$, $L_{c}=20\,{\rm nm}$, $m_\star=0.067\, m_e$ with $m_e$ the bare electron mass.
		Panel~(b): energy spectrum of the polaritonic modes as a function of the cavity frequency for $n_e=1.1\times 10^{12} \rm cm^{-2}$, corresponding to a Fermi level energy $E_F$ right below the pseudo-continuum [cf. Panel~(a)].
		In this specific configuration the largest polaritonic splitting at resonance corresponds to the $\nnu=\{3,1\}$ transition, with $\Omega_\nnu^{\rm res}/\omega_{\nnu}\simeq14\%$. 
		The curve color reflects the electronic (blue) or photonic (red) content of the polaritonic branch.}
\end{figure}

Let us start by studying a single quantum well whose potential barrier is $V_0$ and whose spatial width is $L_{QW}$, located at  the middle of the semiconductor cavity spacer of length $L_c$ [Fig.~\ref{fig:OneWell}(a)]. We set $V_0$ in order to have a single quantum confined state in the quantum well, followed by a quasi-continuum of states which are delocalized over the whole semiconductor region.

In Fig.~\ref{fig:OneWell}(b) we present the polariton spectrum as a function of the cavity frequency $\omega_c$.
In order to maximize the light-matter coupling, we set $E_F=E_2$ [cf. panel~(a)].
We can see that some ``dark'' states are completely uncoupled to the photonic mode, while the others show the typical anti-crossing behavior of polaritonic excitations \cite{WeisbuchPRL92,LiuPRB97,DiniPRL03}.
The region in which the polaritonic excitations manifests a hybrid light-matter nature is depicted in Fig.~\ref{fig:OneWell}(b) via the color of the dispersion curve. The width of this region is proportional to the Rabi frequency $\Omega_\nnu^{\rm res}$.

\begin{figure}
	\includegraphics[width=0.95\linewidth]{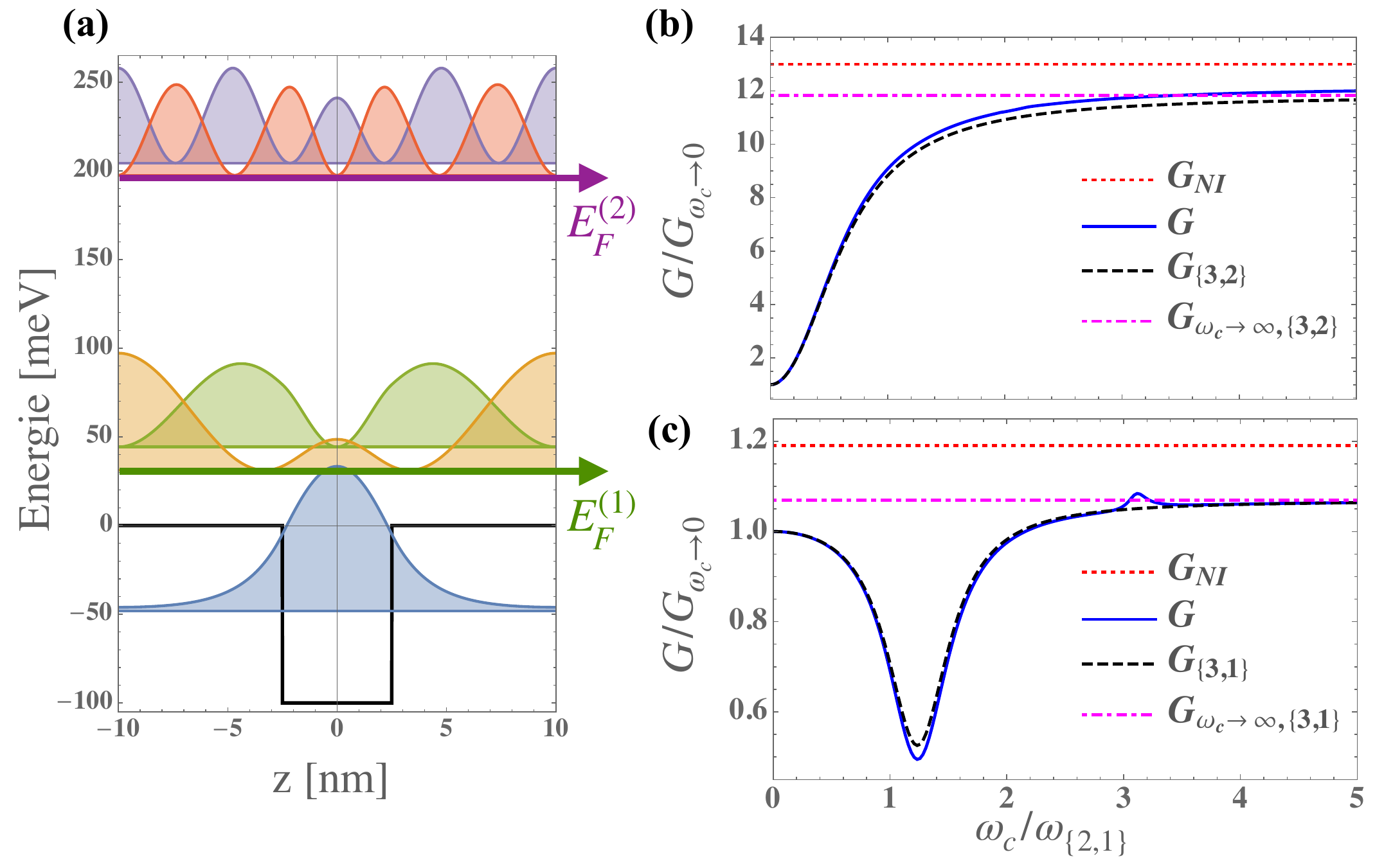}
	\caption{\label{fig:G}
		Panel~(a): same as Fig.~\ref{fig:OneWell}(a).
		We consider here two values $E_F^{(1)}$ and $E_F^{(2)}$ for the Fermi energy level $E_F$.
		Panels~(b,c): cavity conductance $G$  (normalized to $G_{\omega_c\to 0}$)  as a function of the cavity frequency $\omega_c$ (in units of the first transition frequency $\omega_{\{2,1\}}$).
		The blue curves mark the conductance and the red dotted lines the noninteracting limit of the full model including $40$ conduction subbands. The dashed black lines correspond to the cavity conductance calculated within the two-subband approximation and the pink dot-dashed lines mark the analytic asymptotic value $G_{\omega_c\to\infty,\bar{\nnu}}/G_{\bar{\nnu}}$.
		Panel~(c): results obtained for a Fermi energy level  $E_F^{(1)}=E_2$, corresponding to a doping density $n_e=1.1\times 10^{12} \rm cm^{-2}$.
		The dominant transition occurs for $\bar{\nnu}=\{3,1\}$ with $\hbar\omega_{\bar{\nnu}}=92.6\,{\rm meV}$, $\Omega_{\bar{\nnu}}^{\rm res}=0.14\,\omega_{\bar{\nnu}}$, $n_{\bar{\nnu}}=n_e$, and $\Xi_{\bar{\nnu}}^{\bar{\nnu}}=0.05\,\omega_{\bar{\nnu}}$.
		Panel~(b): results obtained for $E_F^{(2)}=E_4$, corresponding to a doping density $n_e=7.9\times 10^{12} \rm cm^{-2}$.
		The dominant transition occurs for $\bar{\nnu}=\{3,2\}$ with $\hbar\omega_{\bar{\nnu}}=13.2\,{\rm meV}$, $\Omega_{\bar{\nnu}}^{\rm res}=1.7\,\omega_{\bar{\nnu}}$, $n_{\bar{\nnu}}=0.18\times10^{12}{\rm cm}^{-2}$, and $\Xi_{\bar{\nnu}}^{\bar{\nnu}}=2.92\,\omega_{\bar{\nnu}}$. The Drude scattering time is $\tau_0=1\,{\rm ps}$.}
\end{figure}

For the same configuration of Fig.~\ref{fig:OneWell}, Fig.~\ref{fig:G}(b) and (c) show the cavity-embedded conductance $G$ normalized to $G_{\omega_c\to 0}$  as a function of the resonator frequency $\omega_c$ in two different configurations for the Fermi energy [see panel~(a)]. 
We first consider an electronic density $n_e$ in such a way that the Fermi energy level is $E_F^{(1)}=E_2$. In this configuration, the confined subband is maximally-populated without filling the delocalized subband. 
The blue curve of Fig.~\ref{fig:G}(c) represents the conductance, exhibiting sharp resonances while changing the cavity mode frequency. These features match the polaritonic resonances observed in Fig.~\ref{fig:OneWell}(b) (coupled $1\to 3$ and $1\to 5$ transitions). At the main resonance, the conductance reduction is approximately $50\%$.
This effect is due to the scattering-time mixing [Eq.~\eqref{Eq:ScatteringTimesMix}].Note that the difference between $G_{\omega_c\to 0}$ and the noninteracting conductance $G_{\!N\!I}$ (the dotted red line) is about a $20\%$ reduction coming from the competition of orbital and energetic effects. In this configuration, the energy subbands are sufficiently spaced for the two-subband approximation of Sec.~\ref{sec:MainTransition} to hold.The dashed black curve of Fig.~\ref{fig:G}(c) has been obtained considering the $1\to3$ transition.
As we can see, the two-subband approximation gives a good qualitative insight about the behavior of the conductance even if we miss the peak corresponding to the $1\to5$ transition. The dot-dashed pink line represents the limit of equation \eqref{Eq:2SBA0}  and shows how the light-matter coupling, even for a far-detuned cavity, renormalizes the conductance compared to the cavityless case. We note that the inequality \eqref{Eq:Inequalities} holds and that the cavity tends to reduce the effects of the depolarization shift with respect to the non-interacting case.

In Fig.~\ref{fig:G}(b), the Fermi Energy $E_F^{(2)}$ was fixed just below $E_4$, in the continuum. Thus further transitions are allowed, giving significant contributions to the conductance. Here the depolarization shift reduces the conductance of almost one order of magnitude  with respect to the non-interacting case. This large effect is smoothly weakened by the light-matter coupling when the cavity frequency is increased. We still can observe residual effects of $2\to n $ resonances, but in this configuration the scattering-time mixing is much less relevant than the orbital and energetic effects.
Remarkably, by increasing the cavity frequency the dark conductance increases by nearly an order of magnitude for the considered range of parameters.
In panel~(b), we also perform the two-subband approximation by restricting to the $2\to3$ transition. Such approximation shows a good agreement, since the dashed black line matches well the blue curve.

	\begin{figure}
		\includegraphics[width=0.95\linewidth]{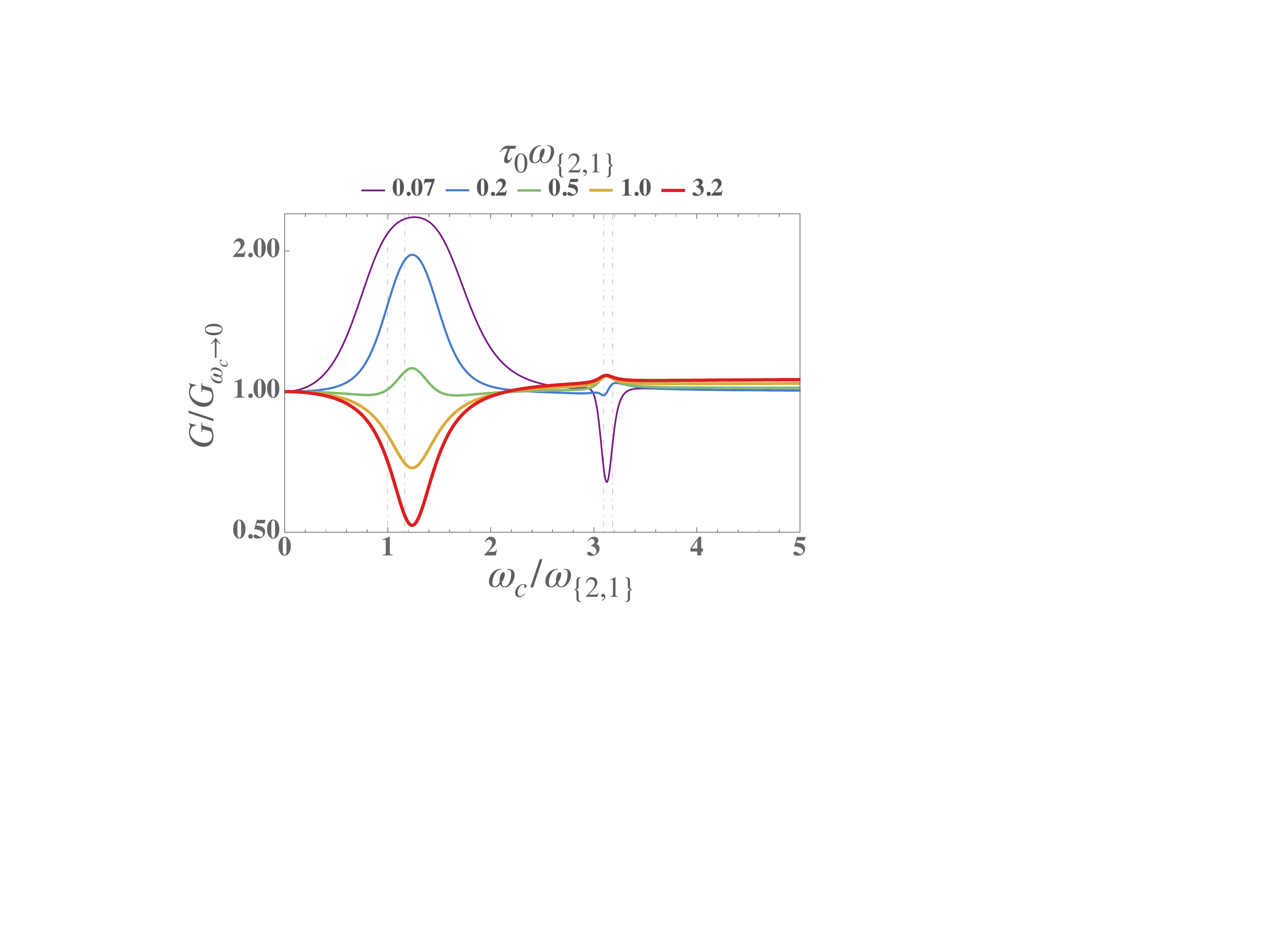}
		\caption{\label{fig:largeTau}
			 Cavity conductance $G$  (on Log scale and normalized to $G_{\omega_c\to 0}$)  as a function of the cavity frequency $\omega_c$ (in units of the first transition frequency $\omega_{\{2,1\}}$).
			Same parameters as in Fig.~\ref{fig:G} when $E_F=E_F^{(1)}$. The different curves correspond to different values of the scattering time $\tau_0$ (see legend).
			Convergence for these results have been obtained including 40 subbands.}
		
	\end{figure}
	
	A critical parameter in determining the sign of the cavity-induced effects is the value of the Drude transport scattering time $\tau_0$.
	To elucidate its role, we take the configuration of  Fig.~\ref{fig:OneWell} and show in Fig.~\ref{fig:largeTau} the ratio $G/G_{\omega_c\to 0}$ for different values of $\tau_0$.
	When we scan over the cavity frequency, the scattering times $\tau_r$ get a mixed light-matter nature around $\omega_c\simeq\omega_\nnu$ and, for $\tau_p\gg\tau_0$, increases compared to the noninteracting case [see Eq.~\eqref{Eq:ScatteringTimesMix}].
	According to Eq.~\eqref{eq:GInt}, if $\tau_0\omega_\nnu\gg 1$ the subband $\nnu$ gives a contribution to the conductance $\propto 1/\tau_r$, so that the conductance due to this transition decreases with respect to the cavity-less case.
	On the contrary when $\tau_0\omega_\nnu\ll 1$, the conductance is proportional to $\tau_r$, which leads to an enhancement of the contribution of the $\nnu$ transition.

	\subsection{Multiple Quantum Wells}
	
	\begin{figure*}
		\includegraphics[width=0.8\linewidth]{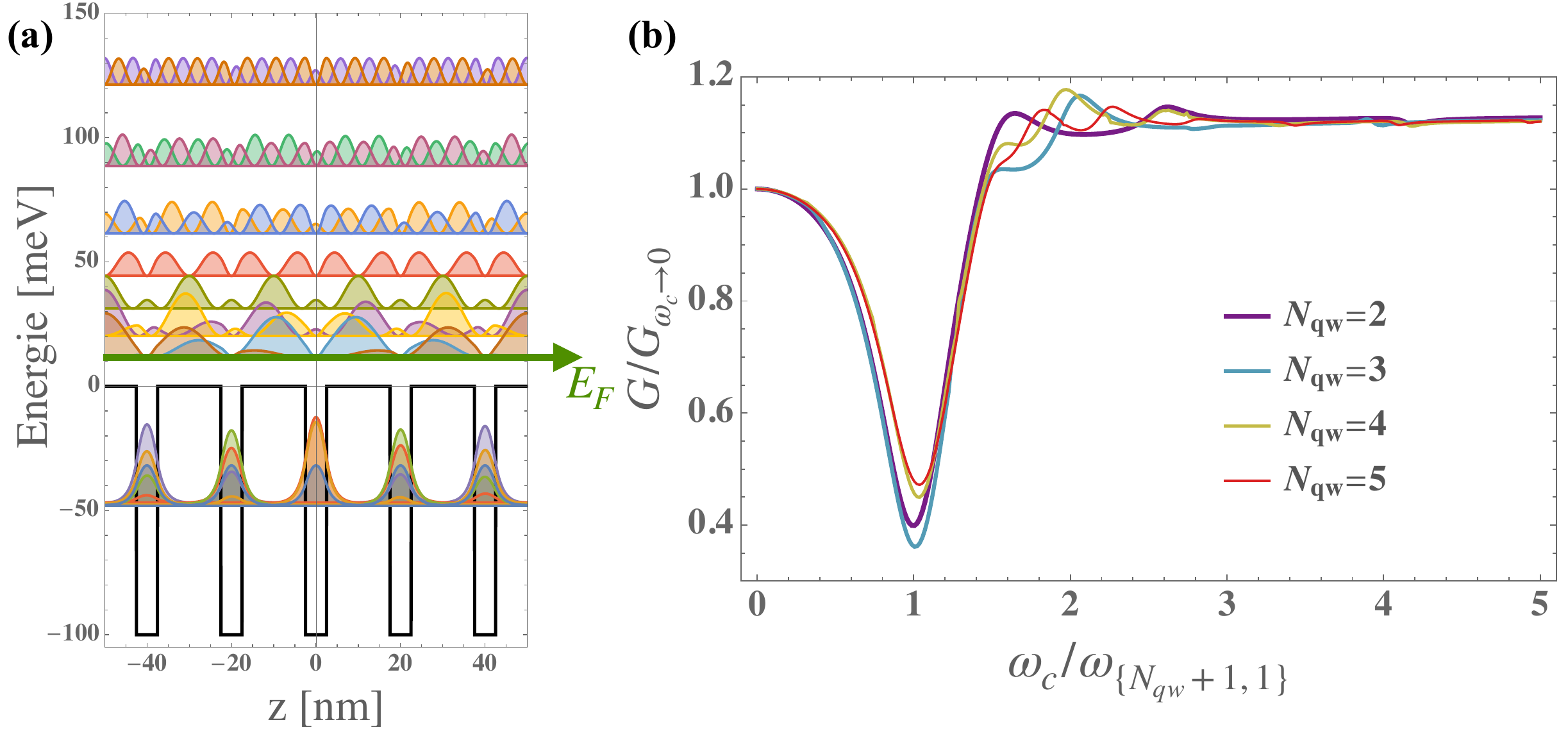}
		\caption{
			\label{fig:Multi}
		Panel~(a): single-electron energy levels and wavefunctions (square modulus, arbitrary units, offset vertically for clarity) for a cavity-embedded heterostructure with 5 identical quantum wells.
			System parameters: $V_0=100\, {\rm meV}$, $L_{QW}=5\,{\rm nm}$, $L_{c}=100\,{\rm nm}$, $m_\star=0.067\, m_e$ with $m_e$ the bare electron mass.
			Panel~(b): normalized cavity conductance  as a function of $\omega_c/\omega_{\{N_{qw}\,+\,1,1\}}$.  The curves correspond to $N_{qw}=2,\,3,\,4,\,5$ (see legend), where $N_{qw}$ is the number of equally-spaced quantum wells of width $L_{qw}=5$, nm giving a total  cavity length $L_c = N_{qw} \times 20$ nm. The other system parameters are unchanged with respect to the case described in panel~(a). The Fermi energy is always fixed just below the energy of the first delocalized subband in order to maximize the interactions. The Drude scattering time is $\tau_0=1\,{\rm ps}$ and  we obtained these results including 40 subbands (necessary to have accurate convergence).}
		
	\end{figure*}
	
In view of experimental realizations, it can be more practical to consider larger cavity spacers. Increasing the cavity length decreases the light-matter interaction [cf. Eq.~\eqref{Eq:Omega}].
However, this can be compensated by considering multiple quantum wells.
For example, in Fig.~\ref{fig:Multi}(a) we represent the single-particle energy levels of a multiple quantum well heterostructure with a number $N_{qw}=5$ of quantum wells. In Fig.~\ref{fig:Multi}(b) we consider different values of $N_{qw}$: for each case, we consider a cavity length $L_c = N_{qw} \times 20$ nm and the Fermi energy is adjusted just below the quasi-continuum. All the confined subbands are approximately equally populated since the tunneling coupling between the different quantum wells is negligible.
The results  in  Fig.~\ref{fig:Multi}(b) show the  normalized conductance as a function of the cavity frequency for the different values of $N_{qw}$ that we have considered. It is apparent that we observe a similar behavior while changing $N_{qw}$. Indeed, the collective coupling due to a succession of quantum wells approximately compensate the decrease of coupling due to the increase of the cavity length \cite{DickePhysRev1954,BonifacioPRA1970,BonifacioPRA1971,VasanelliPRL2015}.

	\section{Conclusions}
	\label{sec:conclusions}
	
	In this work we have introduced a linear-response manybody formalism for the vertical conductance of an arbitrary n-doped parabolic semiconductor heterostructure.
	Within our formalism, it is possible to account for the effect of the light-matter coupling when such an heterostructure is embedded in a single-mode electromagnetic resonator in absence of illumination.
	We diagonalized the many-body Hamiltonian in the dipole gauge, accounting for the collective Coulomb interaction (depolarization shift) and the light-matter coupling in a bosonized formalism.
	As summarized by Eq.~\eqref{Eq:DefCurrentVariation} and Eq.~\eqref{Eq:DefChi}, the conductance is controlled by virtual transitions from the manybody ground state to the excited states via the collective current operator $\hat{J}^\nabla_z$. The conductance depends on the matrix elements of the current operator between ground and polariton excited states, on the energy of such polariton excitations, and on the scattering times.
	
	After presenting our general bosonized theory, we focused on the paradigmatic example of cavity-embedded quantum well heterostructures.
	We have shown that the cavity dark vertical conductance (no illumination) can be largely modified. This effect can be measured by observing the dependence of the conductance on the cavity mode frequency for a fixed cavity length $L_c$. The hybridization of the scattering times leads to the appearance of resonances in the conductance as a function of the cavity frequency.
	These peaks can correspond to enhancement or suppression of the conductance depending on the value of the Drude electronic transport scattering time. Then we focused on the case of multiple quantum wells, showing that a similar phenomenology can be observed by using larger cavity lengths.
	Using parameters for $GaAs$-based semiconductor heterostructures, we have shown both qualitative effects (appearance of a multi-peak/dip structure) and quantitative effects (large enhancement or suppression).
	For a given quantum well structure, the largest vacuum effects are obtained when  the Fermi energy lies above the quantum well barrier.
	As possible future developments of the present theory, we mention the  case of dispersive multimode cavities and the investigation of the regime where the number of electrons is not large enough to allow for a bosonic treatment of the elementary excitations.
	
	Our findings show the role of electromagnetic-vacuum effects on the vertical electronic transport of cavity-embedded semiconductor heterostructure. 	The flexibility of the formalism allows to generalize our theory on any kind of heterostructure, which is all the more significant since several devices are based on this technology. 
	Among the possible generalizations, we mention the extension to systems with non-parabolic dispersions and band wave-mixing.
	
	\section{Acknowledgments}
	
	We thank A.D. Stone, R. Colombelli, J. Faist, C. Sirtori, G. Scalari, Y. Todorov for discussions.
	
	\appendix
	
	\section{Hopfield-Bogoliubov matrix for the considered system}\label{App:Hopfield}
	The Hopfield-Bogoliubov matrix $\bf M$ associated with the Hamiltonian \eqref{Eq:TotalHamiltonian} can be written in the following block form:
	\begin{equation}\label{Eq:App:M}
	{\bf M}=\begin{pmatrix} 
	\omega_c      &     -O     &   0         &  O \\
	-O ^\dagger     &   W+D    & -O ^\dagger  &  -D \\
	0          &   O     &   -\omega_c        &   -O \\
	-O ^\dagger    &    D     & -O ^\dagger  &  -W-D\\
	\end{pmatrix}
	\end{equation}
	with $O=(i\Omega_{\nnu_1},i\Omega_{\nnu_2},\cdots)$, $W=\begin{pmatrix}\omega_{\nnu_1} & & \\ & \omega_{\nnu_2} & \\ & & \ddots\end{pmatrix}$ and $D=\begin{pmatrix}2\Xi_{\nnu_1}^{\nnu_1}&2\Xi_{\nnu_1}^{\nnu_2} &\cdots  \\ 2 \Xi_{\nnu_2}^{\nnu_1}  & 2\Xi_{\nnu_2}^{\nnu_2} & \cdots \\ \vdots&\vdots &\ddots  \\ \end{pmatrix}.$
	
	When restricting to a single main transition of index $\bar{\nnu}$, ${\bf M}$ is a four by four matrix:
	\begin{equation}\label{Eq:App:Mreduced}
	{\bf M}=\begin{pmatrix} 
	\omega_c      &     -i\Omega_{\bar{\nnu}}     &   0         &  i\Omega_{\bar{\nnu}}  \\
	i\Omega_{\bar{\nnu}}      &   \omega_{\bar{\nnu}}+2\Xi_{\bar{\nnu}}^{\bar{\nnu}}    & i\Omega_{\bar{\nnu}}   &  -2\Xi_{\bar{\nnu}}^{\bar{\nnu}} \\
	0          &   i\Omega_{\bar{\nnu}}      &   -\omega_c        &   -i\Omega_{\bar{\nnu}}  \\
	i\Omega_{\bar{\nnu}}     &    2\Xi_{\bar{\nnu}}^{\bar{\nnu}}     & i\Omega_{\bar{\nnu}}   &  -\omega_{\bar{\nnu}}-2\Xi_{\bar{\nnu}}^{\bar{\nnu}}\\
	\end{pmatrix}.
	\end{equation}

	\section{Details about the linear-response formalism for the vertical nonlocal conductivity}\label{App:Kubo}
	
	In this Appendix we wish to show some intermediate steps of the derivation of Eqs.~\eqref{Eq:DefCurrentVariation}, \eqref{Eq:DefChi}, and~\eqref{Eq:CurrentOperatorReduced} from a general linear-response function formalism.
	In the absence of external magnetic field and for an applied electric field $E_z$ applied along the growth direction $z$, the current density variation in the sample is given by~\cite{FlensbergBook}
	\begin{equation}\label{Eq:DefCurrentVariationGeneral}
	\delta\langle\hat{J}_z({\bf r})\rangle=\int\!\! d^3{\bf r}'\, \widetilde{\chi}({\bf r},{\bf r}')\,E_z({\bf r}').
	\end{equation}
	The general nonlocal current-current response function $\widetilde{\chi}({\bf r},{\bf r}')$ reads
	\begin{align}\label{Eq:DefChiGeneral}
	\widetilde{\chi}({\bf r},{\bf r}')=&2\hbar\sum_{exc}\frac{\eta_{exc}}{{\mathcal E}_{exc}-{\mathcal E}_{\rm GS}}
	\nonumber\\&
	\frac{
		\braket{{\rm GS}|\hat{J}^\nabla_z({\bf r})|exc}
		\!\!
		\braket{exc|\hat{J}^\nabla_z({\bf r}')|{\rm GS}}
	}{\left({\mathcal E}_{exc}-{\mathcal E}_{\rm GS}\right)^2+\eta_{exc}^2},
	\end{align}
	where $\hat{J}_z^ \nabla(\boldsymbol{r})$ is the $z$-component of the paramagnetic current operator, namely
	\begin{align}
	\hat{J}_z^ \nabla(\boldsymbol{r})=-i \frac{ e\hbar}{2m_\star} 
	\left[
	\left(\partial_z\hat{\psi}^\dagger\!(\boldsymbol{r})\right)\hat{\psi}(\boldsymbol{r})
	-\hat{\psi}^\dagger\!(\boldsymbol{r}) \left({\partial_z}\hat{\psi}(\boldsymbol{r})\right)
	\right].
	\end{align}
	In the formalism of second quantization on the basis of non-interacting one-electron wave functions~\eqref{Eq:WaveFunctions}, the current density operator reads
	\begin{align}\label{Eq:CurrentOperatorTotal}
	\hat{J}_z^\nabla(\boldsymbol{r})=-i \frac{ e\hbar}{2m_\star}\sum_{i,j,{\bf k},{\bf q}}\xi_{i,j}(z)
	\frac{e^{i {\bf q}\cdot\boldsymbol{r}_\parallel}}{S}\hat{c}_{i,{\bf k}}^\dagger\hat{c}_{j,{\bf k}-{\bf q}}.
	\end{align}
	Since the heterostructure is translationally invariant in the $xy$-plane, the application of a bias voltage along $z$ can only create a field $E_z({\bf r})=E_z(z)$. Similarly, the current variation $\delta\langle\hat{J}_z({\bf r})\rangle$ can not depend on ${\bf r}_\parallel$.
	We can thus integrate over the transverse surface $S$ both sides of Eq.~\eqref{Eq:DefCurrentVariationGeneral} and get
	\begin{equation}\label{Eq:DefCurrentVariationIntermediate}
	\delta\langle\hat{J}_z(z)\rangle=
	\int \!\! dz'
	\left[\frac{1}{S}
	\int\!\! d^2{\bf r}_\parallel \int\!\!d^2{\bf r}'_\parallel\, \widetilde{\chi}({\bf r},{\bf r}')
	\right]
	E_z(z').
	\end{equation}
	The only dependence on ${\bf r}$ and ${\bf r}'$ of Eq.~\eqref{Eq:DefChiGeneral} is in the current density operators.
	Hence, by introducing
	\begin{equation}\label{Eq:JzDef}
	\hat{J}_z^ \nabla(z)\equiv \frac{1}{S}\,\int_S\!\! d^2{\bf r}_\parallel \hat{J}_z^ \nabla({\bf r}),
	\end{equation}
	the term in square brackets in Eq.~\eqref{Eq:DefCurrentVariationIntermediate} reads as $\chi(z,z')$ of Eq.~\eqref{Eq:DefChi}, and Eq.~\eqref{Eq:DefCurrentVariationIntermediate} becomes Eq.~\eqref{Eq:DefCurrentVariation}.
	Upon insertion of Eq.~\eqref{Eq:CurrentOperatorTotal} into Eq.~\eqref{Eq:JzDef} one finally gets Eq.~\eqref{Eq:CurrentOperatorReduced}.
	
	\bibliography{bibliography_Multisubband}
	
\end{document}